\documentclass[amsmath,amssymb,aps,pre,reprint, showpacs, superscriptaddress]{revtex4-2}

\usepackage{amsmath,amsfonts,amssymb,amsthm}
\usepackage{enumerate}
\usepackage{graphicx}
\usepackage{hyperref}

\usepackage{xcolor}

\begin{document}

\title{Signal propagation and linear response in the delay Vicsek model}

\author{Daniel Gei\ss} \affiliation{Institute for Theoretical Physics, University of Leipzig, 04103 Leipzig, Germany} 
\affiliation{Max Planck Institute for Mathematics in the Sciences, 04103 Leipzig, Germany}
\author{Klaus Kroy}  \affiliation{Institute for Theoretical Physics, University of Leipzig, 04103 Leipzig, Germany}
\author{Viktor Holubec} \affiliation{Faculty of Mathematics and Physics, Charles University, CZ-180 00 Prague, Czech Republic}
\email{viktor.holubec@mff.cuni.cz}

\begin{abstract}
Retardation between sensation and action is an inherent biological trait.  Here we study its effect in the Vicsek model, which is a paradigmatic swarm model. We find that: (i) a discrete time delay in the orientational  interactions  diminishes the ability of strongly
aligned swarms to follow a leader and, in return, increases their stability against random orientation fluctuations; (ii) both longer delays and higher speeds favor ballistic over diffusive spreading of information (orientation) through the swarm; (iii) for short delays, the mean change in the total orientation (the order parameter) scales linearly in a small \textcolor{black}{orientational bias of the leaders} and inversely in the delay time, while its variance first increases and then saturates with increasing delays; (iv) the linear response breaks down when orientation conservation is broken.
\end{abstract}

\maketitle

%

\section{Introduction}

Information exchange in animal groups has been suggested to foster survival by improving foraging~\cite{pitcher1982fish} and predator avoidance~\cite{neill1974experiments,milinski1978influence,caraco1980avian,heppner1997three,michaelsen2002magic}. The emerging collective behavior is associated with \textit{swarm intelligence}~\cite{bonabeau1999swarm,dorigo2007swarm} and a \textit{collective mind}~\cite{couzin2007collective}. 
Indeed, if short-range interactions between group members give rise to scale-free correlations~\cite{cavagna2010scale,cavagna2010scale,bialek2014social,vanni2011criticality}, this allows for an efficient information propagation through large groups~\cite{attanasi2015emergence,cavagna2010scale,bain2019dynamic}. And their sensory range is thereby increased way beyond that of its individual members. Inspired by a historical naval battle, where peer-to-peer signalling was employed to  spy beyond the horizon, this is sometimes referred to as the \textit{Trafalgar effect}~\cite{treherne1981group,caro2005antipredator}. It is plausible that animal interactions were evolutionary optimized: too low mutual sensitivity might jeopardize a group's cohesion, too high sensitivity could magnify random perturbations into involuntary collective spasms. As examples for advanced collective strategies, previous work has identified  leadership~\cite{chase1974models,nagy2010hierarchical,shen2008cucker,balazs2020adaptive}, collective escape waves~\cite{herbert2015initiation,procaccini2011propagating}, and hierarchical structures~\cite{chase1974models,nagy2010hierarchical,shen2008cucker,balazs2020adaptive}. From a physical perspective, studies of motile ensembles,
ranging from bacterial colonies~\cite{zhang2010collective,ben1994generic} to flocks of birds~\cite{cavagna2018physics,procaccini2011propagating}, are often subsumed into the field of \textit{active matter}~\cite{gompper20202020,bechinger2016active,ramaswamy2010mechanics}. The associated (active) many-body-theory perspective has already helped to uncover a variety of exotic non-equilibrium collective phenomena~\cite{cates2015motility,hagan2016emergent,bauerle2018self,farrell2012pattern,solon2015pattern} and hinted at potential technological applications~\cite{brambilla2013swarm,vernerey2019biological,khadka2018active,muinos2021reinforcement}.

A key ingredient in the interactions of living agents is a time delay between perception and reaction. Signal propagation and processing, as well as actuation suffer from inevitable speed limitations. In general, the associated time delay in the mutual interactions between the agents may enrich the resulting collective dynamics with a hierarchy of instabilities, including oscillations, multistability, and  chaos~\cite{geiss2019brownian,foss1996multistability,wernecke2019chaos}. These have been the subject of intense study in the field of dynamical systems \cite{atay2010complex}. Conversely, delays have also been employed to stabilize unstable orbits in chaotic systems, e.g., via the Pyragas method~\cite{pyragas1995control}. 
In the context of information propagation, time delay has moreover been investigated in connection with sensory~\cite{Peng2006}, social~\cite{Yin2021} and financial  (bitcoin)~\cite{decker2013} networks, as well as in human traffic~\cite{Baccelli2012}. For self-propelling particles with retarded  pairwise attractions, time delays can trigger behavioral patterns~\cite{forgoston2008delay,mier2012coherent} reminiscent of dynamical regimes known from (linear) stochastic delay differential equations~\cite{geiss2019brownian}. Small perception-reaction delays in a \textit{Vicsek model} (VM) may foster clustering~\cite{piwowarczyk2019influence} and flocking~\cite{holubec2021scaling}, similarly as in the Cucker--Smale model~\cite{Erban2016}. In contrast, long delays have the opposite effect~\cite{Erban2016,piwowarczyk2019influence,holubec2021scaling}. From a more technological perspective, time delays naturally play an important role in feedback systems~\cite{khadka2018active,muinos2021reinforcement}. \textcolor{black}{Finally, it was recently suggested that retardation induced by inertia~\cite{attanasi2014information,attanasi2015emergence,lowen2020inertial} or by discrete time lags~\cite{holubec2021scaling} could be the missing link between the VM and the behavior empirically  observed in natural swarms~\cite{Cavagna2017}. In this respect, 
it was moreover found that the inertial version of the VM (the so-called inertia spin model) better describes collective turns as observed in natural swarms than the standard VM~\cite{cavagna2015flocking}. However, to the best of our knowledge, there are no other studies of delay effects on the information propagation and response of motile active matter systems.}

In the following, we \textcolor{black}{therefore} study retarded information spreading and linear response in the two-dimensional delay VM, defined in Sec.~\ref{Sec VM}. First, we focus on the ability of an aligned system (flock) to perform collective turns (Sec.~\ref{Sec Turns}), and its reaction to local perturbations~\cite{Geiss2021lattice} (Sec~\ref{Perturbation}). For the former case,
we study in Sec.~\ref{Sec Signal} how the signal spreads in space and time and determine the corresponding dispersion relation. In Sec.~\ref{sec: linear response}, we consider the response of the system to an orientational bias applied to a subgroup of agents (so-called leaders) and in particular evaluate the conditions for linearity~\cite{chate2008collective,pearce2016linear}. Finally, Sec.~\ref{Sec: Conclusion} summarizes our findings and gives an outlook to possible extensions of the present work. 

\section{Time delay Vicsek model}\label{Sec VM}

The VM~\cite{vicsek1995novel,chate2008modeling} is a paradigmatic model for dry active matter~\cite{ramaswamy2010mechanics}. Its original version~\cite{vicsek1995novel} consists of $N$ mobile spins moving with constant speed $v_0$, and interacting via alignment interactions. We consider the two-dimensional version with time delayed metric interactions~\cite{holubec2021scaling}, where each particle at time $t+1$ assumes the average orientation of all neighbors at distance less than $R$ at an earlier time $t-\tau < t$, up to some noise. Velocities $\mathbf{v}_i(t)$ and positions $\mathbf{r}_i(t)$ of the individual particles obey the equations of motion
\begin{align}
\mathbf{v}_i(t+1) &= v_0 \mathcal{R}_\eta \left[
\mathbf{v}_i(t) + \mbox{$\sum_{j\neq i }$} n_{ij}(t-\tau) \mathbf{v}_j(t - \tau) 
\right],
\label{eq:vtdisc}\\
\mathbf{r}_i(t+1) &= \mathbf{r}_i(t) + \mathbf{v}_i(t+1),
\label{eq:rtdisc}
\end{align}
where the noise operator $\mathcal{R}_\eta\mathbf{v}$ realizes a uniform random rotation by an angle in the range $[-\eta,\eta]$ around its  normalized argument $\mathbf{v}/|\mathbf{v}|$. For the chosen metric interactions, the elements of the connectivity matrix $n_{ij}$ are given by $n_{ij}(t) = 1$ if $r_{ij}(t) = |\mathbf{r}_i(t) - \mathbf{r}_j(t)| < R$ and $n_{ij}(t) = 0$ otherwise. We take $R$ as our length unit, i.e., we set $R=1$, and the time step $\Delta t=1$  between two consecutive interactions as the time unit. In contrast to the closely related XY model~\cite{kosterlitz1974critical} the connectivity matrix is time-dependent. In the VM with topological interactions, which seems more appropriate for natural bird flocks~\cite{ballerini2008interaction}, we expect qualitatively similar results for information spreading as described below~\cite{Geiss2021lattice}.

Without a time delay, the model~\eqref{eq:vtdisc}--\eqref{eq:rtdisc} is just the standard VM \cite{vicsek1995novel}, which exhibits  overdamped orientational dynamics of $\mathbf{v}_i(t)$~\cite{Cavagna2017,holubec2021scaling}. At (imposed) high densities, a highly aligned phase emerges, reminiscent of flocking birds. The disordered state found at low densities resembles an insect swarm. \textcolor{black}{For large particle numbers,} the transition between these two phases is discontinuous (`first order')~\cite{ginelli2016physics}. A finite delay time $\tau$~\cite{holubec2021scaling} or, alternatively, some finite  inertia~\cite{cavagna2021equilibrium} in the dynamical equation for the velocity is known to bring the model predictions closer to the space-time correlations and finite-size scaling observed in natural swarms~\cite{Cavagna2017}. \textcolor{black}{Moreover, inertia was argued to stabilize the flock phase and improve the ability of the system to perform collective turns~\cite{cavagna2015flocking}. In the next section, we show that time delay in the interactions, which is often likened to inertia, yields different effects.}

\section{Collective maneuver} \label{Sec Turns}

To sustain their cohesion, flocks \textcolor{black}{need to be} able to perform collective maneuvers~\cite{cavagna2015flocking}. We now investigate the ability of individual members to initiate such maneuvers in the delay VM. Specifically, we consider an initially ($t=0$) fully polarized flock, with all particles moving in the same direction $\mathbf{v}_{\text{init}}/v_0=(1,0)^\intercal$, corresponding to the  polarization $\Phi=(\sum_i \mathbf{v}_i)^2/v_0N = 1$. The initial positions are distributed randomly inside a circle of diameter $L_\text{F}$ around the origin. The only exception is a single \textit{leader} placed at the front of the flock, $\mathbf{r}_{\text{L}}(0)=(L_\text{F},0)^\intercal$. To initialize the particle velocities, we assume that the whole flock had evolved along the $x$-axis according to Eqs.~\eqref{eq:vtdisc}, \eqref{eq:rtdisc} for  $t<0$, while, for $t\ge 0$, the leader is constrained to make a deterministic turn
\begin{equation}
\mathbf{v}_\mathrm{L}(t+1) = \mathcal{T}_\varphi \mathbf{v}_\mathrm{L}(t)\,.
\label{eq:deterministicL}
\end{equation}
The operator $\mathcal{T}_\varphi$ rotates its argument by an angle $\varphi$. Its deterministic time evolution acts as a persistent source of information fed to the flock dynamics. For the sake of \textcolor{black}{clarity}, we neglect the noise, $\eta(t)=0$, \textcolor{black}{for now. Then the only perturbation in the system is the deterministic motion of the leader, and the initially fully polarized flock can be understood as a steady state configuration of the system.}

\begin{figure}
\centering
\includegraphics[width=1.05\linewidth]{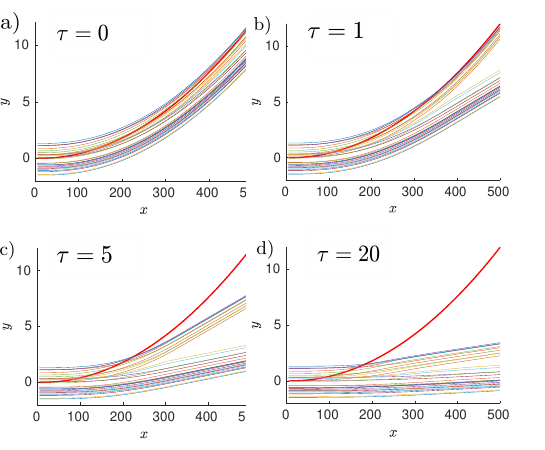}
	\caption{Fate of a flock in the delay Vicsek model. The flock starts with identical initial conditions near the lower left end of each graph \textcolor{black}{and performs $10^3$ time steps} with four different values of time delay ($\tau=0,1,5,20$) in the mutual interactions. The bold red line shows the deterministic turn~\eqref{eq:deterministicL} performed by a leader. With increasing delay, the impact of the leader on --- and its dwell time in --- the flock diminish. The flock partly follows its turn, but looses coherence and may even break up. Parameters used: $N=30,\ L_\mathrm{F}=3,\ v_0=0.5,\ \varphi=5\cdot 10^{-5}$.}	
	\label{fig:trajectories1}
\end{figure}


\textcolor{black}{In Fig.~\ref{fig:trajectories1}, we show typical turning events for four different values of the time delay of a small flock of $N=30$ agents following a slowly turning leader.} For short delays, all the particles turn with the same radius, in a manner that is loosely reminiscent of natural bird flocks~\cite{pomeroy1992structure,attanasi2015emergence,cavagna2015flocking}. However, due to the accumulating mismatch between the navigation of the flock and its leader, the latter ultimately escapes its followers \textcolor{black}{for arbitrary nonzero turning speeds $\varphi$}. Its cumulative impact on the overall flock orientation (the order parameter) is determined by its dwell time \textcolor{black}{in the flock}. Since the conductive spreading of orientation is diffusive~\cite{Geiss2021lattice} \textcolor{black}{and thus inefficient}, the particles that spend most time in the vicinity of the leader follow it more closely than the more remote ones. This leads to a reshuffling of the flock formation and an associated defocusing that can eventually rip the flock apart. Interestingly, the effect becomes increasingly pronounced with growing delay \textcolor{black}{$\tau$}, but fades again beyond a certain threshold \textcolor{black}{$\tau \gtrapprox L_{\rm F}/v_0$}. This non-monotonicity indicates
a trade-off between the delay-induced increase with $\tau$ in the time spent in the vicinity of the leader needed for successful alignment, which favors flock break-up, and a correspondingly shortened dwell time of the leader inside the flock, which reduces its cumulative impact. \textcolor{black}{Taking into account noise, one finds that the splitting events become more frequent with increasing noise amplitude. This is expected because randomly turning particles can act as additional `leaders' initiating new splitting events. Similarly, increasing the particle number $N$ for a given $v_0$ while keeping the initial flock density $N/L_{\rm F}^2$ constant eventually always induces splitting of the flock as the information reaching distant parts of the flock during the finite duration of the turn gets strongly damped. On the other hand, increasing the particle number $N$ while keeping the initial flock size $L_{\rm F}$ constant (thus increasing the initial swarm density) stabilizes the flock (splitting gets less likely) as the relative weight of the individual perturbations decreases. For the same
reason, the ability of the flock to follow a single leader decreases with increasing density.} 

\begin{figure}
\centering
\includegraphics[width=0.7\linewidth]{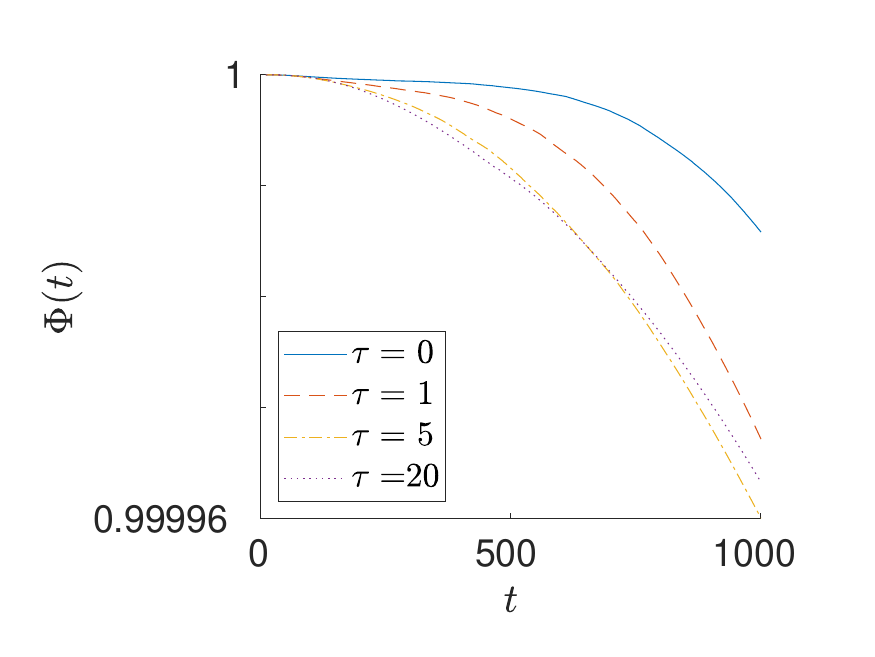}
\caption{Time evolution of the polarization for the flocks depicted in Fig.~\ref{fig:trajectories1}, averaged over 100 simulation runs.}
\label{fig:trajectoriesPol}
\end{figure}

\textcolor{black}{In summary, in} the VM, flocks thus only follow slowly turning leaders. Due to the metric interaction rule, better coherence is naturally achieved by more compact flocks. However, no matter the conditions, the leader eventually always leaves a finite flock, as witnessed by the decreasing polarization in Fig.~\ref{fig:trajectoriesPol}.  The generally poor take-up of the leader's direction by the VM flocks is a direct consequence of the averaging interactions, which disperse the information, preventing a perfect alignment.
Recent investigations~\cite{cavagna2015flocking} show that in the VM augmented with orientational inertia, the turning information can propagate much more efficiently \textcolor{black}{due to an additional degree of freedom that (approximately) conserves the flock's curvature.} As a result, the inertia spin model allows for coherent collective turns where the leader stays inside the flock. Contrary to \textcolor{black}{these findings}, the time delay \textcolor{black}{renders} the ability of the VM to perform coherent turns even worse. This striking difference between the effects of inertia and delay is interesting because other aspects of the two models, e.g., dynamical critical exponent and shape of time correlation functions, are similar~\cite{cavagna2018physics,holubec2021scaling}.

In Sec.~\ref{sec: linear response}, the turning scenario is revisited from the perspective of the flock's linear response. Before that, we discuss another interpretation of the turning event, in which the `poor obedience' of the flock has a more advantageous connotation.

\section{Perturbation} \label{Perturbation}

\textcolor{black}{If the leader makes a sudden (or very fast) turn in the scenario of the preceding section, it can be interpreted as a random perturbation in the flock rather than an intended systematic maneuver.} In stable systems, such perturbations should not create a strong response. This again reflects the necessity for tuning the information propagation in \textcolor{black}{a flock} to an intermediate strength. And, again, delayed interactions can play a role in this respect. Efficient information propagation promotes cohesion and a coherent response of the flock to a stimulus but also tends to make the system overly sensitive to accidental local perturbations that do not qualify as reasonable cues for the whole flock. 

\begin{figure}
    \centering
    \includegraphics[width=1.05\linewidth]{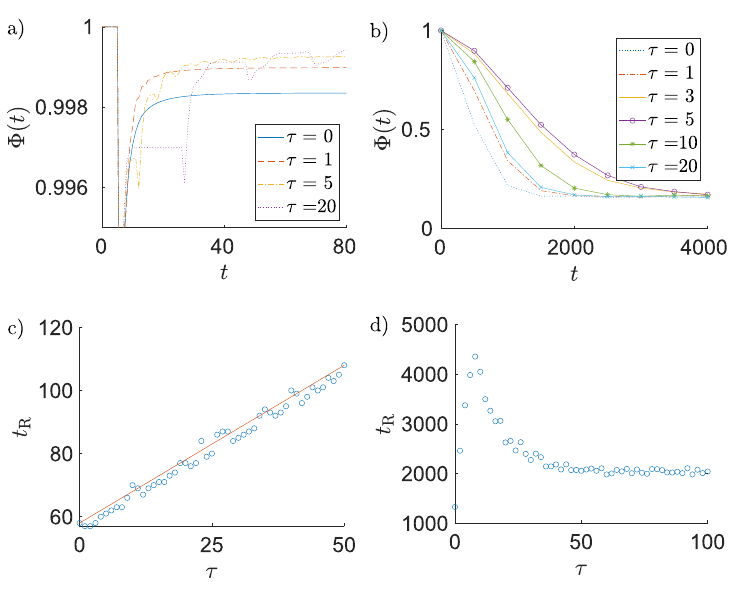}
	\caption{a) Relaxation of the flock's polarization after a sudden change of the leaders orientation by the angle $\varphi=1$ for four different time delays. b) \textcolor{black}{Decay} of the flock's polarization to its disordered stationary value for \textcolor{black}{a strong noise,} $\eta=0.1$. \textcolor{black}{Symbols in panels c) and d) show relaxation times corresponding to a) and b), respectively, as functions of the time delay. The solid line in c) depicts the theoretical expectation $t_{\rm R}(\tau) = t_{\rm R}(0) + \tau$.} Parameters used: $N=30,\ L_\mathrm{F}=6,\ v_0=0.5,\ N_\mathrm{runs}=10^3$.}
	\label{fig:polarizationvarytau}	
\end{figure}

To study the response of the delay VM to such `accidental perturbations', we consider the same \textcolor{black}{noiseless} scenario as in the preceding section --- with the important difference that the leader abruptly turns by an angle $\varphi$ at time $t=0$ but subsequently obeys the dynamic equations~\eqref{eq:vtdisc}, \eqref{eq:rtdisc}, rather than following an externally prescribed route. In this case, the system receives information only at time $t=0$ and relaxes freely, thereafter. The response of the flock in terms of its polarization is plotted in Fig.~\ref{fig:polarizationvarytau}a, again for four different values of time delay. Two main effects of the time delay on the response can be discerned. First, as already gleaned from the above discussion, delayed interactions cause a lag of the collective response. \textcolor{black}{This increases the relaxation time $t_{\rm R}$ of the polarization by the delay time $\tau$, i.e., $t_{\rm R}(\tau) = t_{\rm R}(0) + \tau$. This ansatz is verified in Fig.~\ref{fig:polarizationvarytau}c, where we show the relaxation time of the polarization as a function of the delay time. Here, the relaxation time $t_{\rm R}$ is defined as the time when the polarization reaches 99\% of its plateau value, i.e., $\Phi(t_{\rm R}) = 0.99\, \Phi(\infty)$. Second, a sufficiently strong retardation of the interactions introduces an oscillatory response in the form of periodic spikes in the polarization dynamics. Their period is given by $\tau + 1$ and their amplitude increases with growing $\tau$ and decays with growing lab time $t$. Such delay-induced oscillations} are well known to be a hallmark of delay dynamical systems~\cite{geiss2019brownian,holubec2021scaling}. In the delay VM, they can be explained as follows. After the leader has been perturbed, it will spread the perturbation to the rest of the flock but also tend to realign itself with its (former) neighbors. The leader may therefore already have realigned with the initial direction of its more distant neighbors once these neighbors are affected by the spreading impulse of the initial perturbation. Their delayed reaction may then echo back to the leader, after yet another delay time. This wave-like oscillatory spreading pattern clearly constitutes a pulse propagation. Apart from the crucial delay interaction, it is tied to the finite extent of the perturbing impulse and the subsequent free relaxation of the leader, so that it could not be observed in the turning scenario studied further above, where the leader's dynamics is at all times determined by Eq.~\eqref{eq:deterministicL}, which provides a persistent information influx preventing the backlash. \textcolor{black}{However, once the oscillations in the polarization are present, they represent a robust, time-local effect that does not vanish in large ensembles of particles, not even after averaging over the noise~\cite{holubec2021scaling}, unless one introduces sufficient dispersion in the reaction-time delays of the individual particles.}


So far, we concentrated on purely deterministic dynamics, for simplicity. But qualitatively similar behavior can also be observed for systems affected by noise. One may think of external noise as consisting of a large ensemble of `accidental perturbations' of the same type as just discussed. They permanently trigger random chains of similar orientational adaptation events as just described, \textcolor{black}{across the whole flock.} Let us now consider an initially fully polarized flock whose polarization \textcolor{black}{decays} (to zero, if the flock is infinitely large) due to \textcolor{black}{persistent} noise. In Figs.~\ref{fig:polarizationvarytau}b, we show that also in this case the time delay has a stabilizing effect on the system, as it in general increases \textcolor{black}{the polarization decay time}. However, \textcolor{black}{the latter} does not grow monotonically in $\tau$. Instead, \textcolor{black}{as shown in Fig.~\ref{fig:polarizationvarytau}d, the decay time $t_{\rm R}$, defined in the standard way as $\Phi(t_{\rm R}) = 1/{\rm e}$,} exhibits a peak for intermediate delays, and then decreases to a finite asymptotic value. We conjecture that this peak results from a competition between the reduced alignment strength and the improved stability against perturbations, caused by the delay. Similar non-monotonic effects in the delay VM were previously reported in Refs.~\cite{piwowarczyk2019influence,holubec2021scaling}.

\section{Signal propagation} \label{Sec Signal}

Recent studies have investigated the information spreading in the standard VM~\cite{brown2020information,cavagna2020vicsek,Geiss2021lattice} and also in the inertia VM~\cite{cavagna2018physics}.
In this vein, we now study how information about the leader's turn propagates through a flock, in the delay VM. In our setting, it is not obvious how to employ the usual information theoretical notions such as entropy or channel capacity~\cite{shannon1948mathematical,cover1999elements}. Therefore, our analysis relies on the concept of \textcolor{black}{maximum response~\cite{Geiss2021lattice} or, more generally, maximum correlation~\cite{cavagna2018physics}.} 

The change of state of a Vicsek flock is described by the accelerations $\mathbf{a}_i(t)=\dot{\mathbf{v}}_i(t)$ or, equivalently, by changes $\dot{\theta}(t)$ of the orientations $\theta_i(t)=\arctan(v_{y,i}(t)/v_{x,i}(t))$ of the individual particles. Following Ref.~\cite{Geiss2021lattice}, we take the turning rate $\dot{\theta}(t)$ as a measure for the strength of the signal arriving at the particle $i$ at time $t$. For our specific setup, this approach is equivalent to the acceleration-correlation based approach applied in Refs.~\cite{cavagna2018physics,cavagna2016spatio}. Similarly as in Sec.~\ref{Sec Turns}, we consider in this section an initially highly polarized flock \textcolor{black}{with $\theta_i(0)=0$} and a leader performing a deterministic turn. However, in contrast to the above scenarios we assume that the leader changes its orientation instantaneously at time $t=0$  by an angle $\varphi$ that remains fixed thereafter. In this setting, the signal source is better localized in space and time than in the case of the continuous turn of Sec.~\ref{Sec Turns}. To minimize boundary effects, we also initialize the leader in the center of the flock instead of its front, \textcolor{black}{and we neglect the noise. For a non-vanishing noise, the system becomes unstable unless it is confined by boundary conditions.}

\subsection{Spatial propagation of a signal}

\begin{figure}
    \centering
    \includegraphics[angle=270,width=1.05\linewidth]{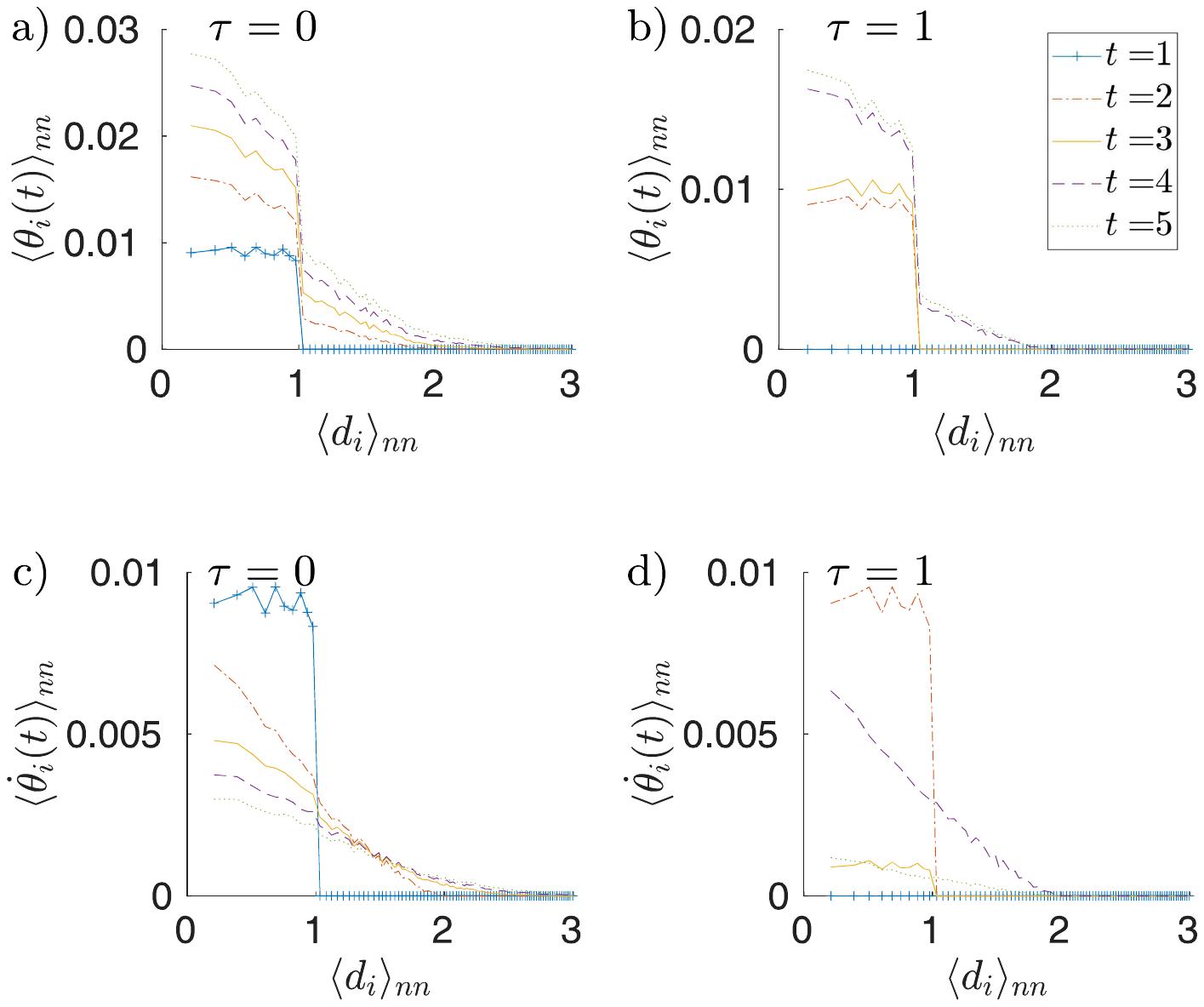}
	\caption{Time evolution of the average orientation $\theta_i(t)$ (a,b) and its rate of change $\dot{\theta}_i(t)$ (c,d) for \textcolor{black}{a stationary flock with} $v_0 = 0$ and $\tau=0,1$.  The signal induced by a leader at the origin spreads in steps of the size of the interaction radius $R = 1$. Delay induces oscillations of $\dot{\theta}_i(t)$ with the period $\tau+1$. Parameters used: $N=1000$, $L_\mathrm{F}=20$, $\varphi=0.1$, and $N_\mathrm{runs}=10^2$.}
	\label{fig:SpatialSignalSpreading2}	
\end{figure}


Let us first discuss the signal spreading through the flock of immobile ($v_0=0$) agents with a static interaction network. In this case, the distance to the source of the signal is a well defined quantity and thus we can take the corresponding results as a standard for further analysis.
Besides, since the lattice is fixed, the information spreads homogeneously and solely by conduction~\cite{Geiss2021lattice}. \textcolor{black}{For a thorough analysis of this scenario for the standard VM ($\tau=0$), see Ref.~\cite{Geiss2021lattice}.}

In Fig.~\ref{fig:SpatialSignalSpreading2}, we show the resulting orientations, $\theta_i(t)$, and their rates of change $\dot{\theta}_i(t)$, as functions of the distance $d_i$ to the leader for $\tau=0$ and $\tau=1$.
To avoid a dependence on the initial condition, the shown values are averaged with respect to $N_\mathrm{runs}$ simulation runs as follows. We first collect all data $\{d_i,\theta_i(t),\dot{\theta}_i(t)\}_{i=1,...,N}$ from the individual simulations runs and sort the total data with respect to the distance $d_i$. The so obtained ordered sets of \textcolor{black}{$N N_\mathrm{runs} $} data points are then smoothened by \textcolor{black}{calculating a sliding average over $N_\mathrm{runs}$ neighboring data points.} The corresponding averages are denoted by the symbol $\left<\bullet \right>_{nn}$.

Panels a) and c) show that, for $\tau=0$, the information spreads in `steps' of the width of one interaction radius $R=1$. After the first time step, the information is homogeneously distributed within the interaction radius $R=1$  because the corresponding particles interacted equally with the \textcolor{black}{localized} information source. \textcolor{black}{The rates $\dot{\theta}_i(t)$ monotonically decrease with the distance from the leader, because particles closer to the leader interact with more neighbors that have already changed their orientation.
For time delay $\tau=1$, the information in panel b) starts spreading only at time $t = \tau + 1 = 2$. Furthermore, temporal oscillations of period $\tau+1$ are induced in $\dot{\theta}_i(t)$, for the same reasons as explained above. In panel d), these oscillations manifest themselves as a large $\dot{\theta}_i(t)$ at times $t=2$ and $4$ and a strongly damped $\dot{\theta}_i(t)$ at times $t=1,3,$ and 5.} Regardless of the time delay, the orientations $\theta_i(t)$ exhibit a discontinuity for $d_i=R$ that increases with time, as a result of the continuous influx of information from the fixed leader.

\begin{figure}
    \centering
    \includegraphics[width=1.05\linewidth]{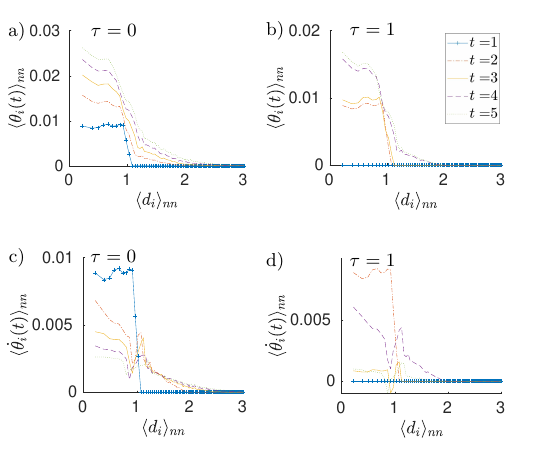}
	\caption{
	Time evolution of the average orientation $\theta_i(t)$ (a,c) and its rate of change $\dot{\theta}_i(t)$ (b,d) for \textcolor{black}{a moving flock with} $v_0 = 0.5$ and $\tau=0,1$. Due to the breaking of the rotational symmetry of the static network \textcolor{black}{(Fig.~\ref{fig:SpatialSignalSpreading2})}, the spreading is enhanced in the direction of the leader's motion. This results in a kink in $\dot{\theta}_i(t)$ at $d_i=R$. Other parameters are the same as in Fig.~\ref{fig:SpatialSignalSpreading2}.}
	\label{fig:SpatialSignalSpreading1}
\end{figure}

For moving flocks with $v_0>0$, the distance traveled by the signal is no longer unambiguously defined. Even though it might seem inadequate, one possible choice of $d_i$ is still the distance to the leader at time $t=0$ as it gives qualitatively the same results as other definitions, e.g., one that is based on a co-moving frame~\cite{Geiss2021lattice}. In what follows, we thus stick to this definition. In Fig.~\ref{fig:SpatialSignalSpreading1}, we report again the orientations, $\theta_i(t)$ and their rates of change $\dot{\theta}_i(t)$, as functions of the (initial) distance $d_i$ to the leader. The resulting curves are similar to those for the static network, as shown in Fig.~\ref{fig:SpatialSignalSpreading2}. However, the discontinuity in $\theta_i(t)$ at the interaction range, $d_i=R$, is now smeared out and the rate $\dot{\theta}_i(t)$ exhibits concomitant kinks. Both effects result from the anisotropy of the \textcolor{black}{`convective'} signal propagation caused by the motion of the leader. For $v_0>0$, both the leader and its 
`interaction zone' move. 
This affects differently particles in the direction of the leader than those in the opposite direction, which we do not distinguish when evaluating the shown results. The downward-kink in $\dot{\theta}_i(t)$ corresponds to the neighbors behind the leader, which leave its interaction zone, and the upward kink corresponds to the particles in front, entering its interaction zone. Averaging these two effects then induces the smearing of the discontinuity of the orientation $\theta_i(t)$ as a function of distance, at the edge of the interaction zone.

\subsection{Dispersion relation}

\begin{figure}
    \centering
    \includegraphics[width=1.05\linewidth]{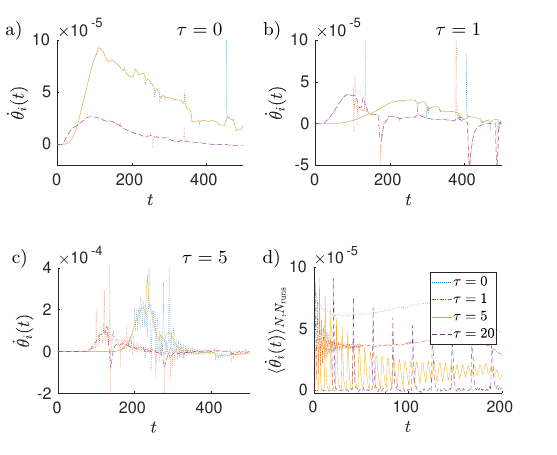}
	\caption{\textcolor{black}{a-c) Time evolution of the signal intensity $\dot{\theta}(t)$, as perceived at two different particles, bare (blue and red dotted lines) and smoothed (solid yellow and dashed purple lines) versions, for $\tau=0,1,5$, and d) averaged over all $N$ particles and $N_\mathrm{runs}$ simulation runs.} Parameters used: $N=10^3,\ L_\mathrm{F}=20,\ v_0=0.5,\ \varphi=0.1$.}	
	\label{fig:ThetaSingle}	
\end{figure}

A central quantity in the study of signal propagation is the \textit{signal speed}, or, equivalently, the \textit{dispersion relation}. However, to define these quantities for the VM is neither trivial nor unique. Consider the typical time evolution of the signal strength, measured in terms of $\dot{\theta}_i(t)$. \textcolor{black}{As shown in Fig.~\ref{fig:ThetaSingle}a, the signal intensity $\theta_i(t)$ reaching an agent $i$ at some time $t>0$ is initially very low and grows over time until it reaches a maximum and starts to decay.} For a \textcolor{black}{stationary or} slowly moving flock, for which the information spreading is dominated by conduction, the time when the signal first reaches a particle at distance $d_i$ from its source can roughly be estimated as $d_i/R$. This is the time required for the signal to arrive from the leader's initial position, if traveling along the shortest path. \textcolor{black}{This initial} signal gets \textcolor{black}{diluted} by a factor $1/N_{\rm int}$ after each time step, since it is distributed among the $N_{\rm int} \approx N R^2/L_\mathrm{F}^2$ nearest neighbors, on average. This explains why it \textcolor{black}{becomes} very weak at its spreading front. As the number of possible paths connecting the initial position of the leader with the nearby particles in the flock increases over time, also their perceived signal intensity, as well as the rate of change $\dot{\theta}_i(t)$ of their orientation, grows. Once the \textcolor{black}{particles} thereby become increasingly aligned with the leader, $\dot{\theta}_i(t)$ decays again. In what follows, we identify the time when $\dot{\theta}_i(t)$ reaches its maximum as the \textcolor{black}{(characteristic)} signal propagation time $T_i$. The corresponding signal speed is $d_i/T_i$. For more details concerning this choice, see Ref.~\cite{Geiss2021lattice}.

To inspect the signal propagation more closely, consider the time evolution of the signal strength at two given particles, as depicted in Fig.~\ref{fig:ThetaSingle}a-c. Each of the three panels corresponds to a different value of the delay. 
The most striking feature of the raw data (dotted blue and red curves) is the presence of sharp spikes, which become more frequent for long time delays. We reckon that these ``resonances'' are due to random inhomogeneities in the initial conditions. At a first glance, identifying the time $T_i$ with these spikes might be feared to lead to erroneous estimates for the signal speed. Therefore, we alternatively considered the smoothed profiles depicted by the solid yellow and dashed purple lines. They were obtained by first performing sliding averages over five raw data points and then substituting the remaining outliers (beyond five times the average $\langle \dot{\theta}_i(t)\rangle_t$ over the shown time interval) with an average value of their neighboring points. But the results for the dispersion relations remained qualitatively the same.

Besides the spurious spikes, the curves for time delays $\tau>0$ again exhibit oscillations, which can even reach negative values as the particle orientations tend to align to their previous values. The period of these oscillations is $\tau+1$ and their magnitude increases with growing delay. For sufficiently long delays, the signal strength exhibits sharp positive and negative peaks as the particle is either aligned with its own initial orientation or with the initial direction of the leader.
These sharp peaks are even better visible in Fig.~\ref{fig:ThetaSingle}d, where we show the average of $\langle \dot{\theta}_i(t) \rangle_{N,N_\mathrm{runs}}$ taken over all $N$ particles and $N_{\rm runs}$ simulation runs, for the same values of time delay as in Fig.~\ref{fig:ThetaSingle}a-c. This quantity measures the total rotation rate in the flock at the given time. For short delays, the signal is initially stronger, but the envelopes of the peaks induced by long delays have the longer decay time.

\begin{figure}
    \centering
    \includegraphics[width=1.05\linewidth]{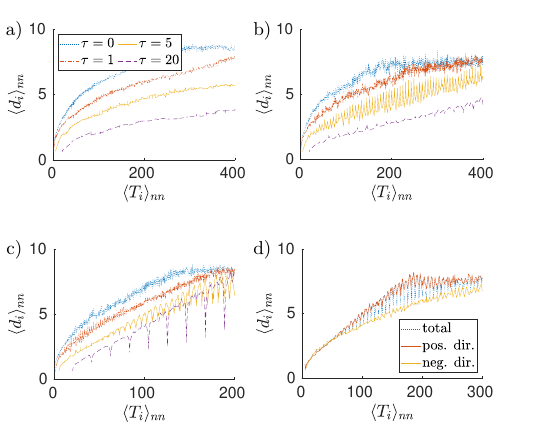}
	\caption{a-c) The dispersion relations obtained as the average distance from the leader $d_i$ at time $T_i$	when the change of the orientation $\dot{\theta}_i(t)$ peaks for $v_0=0$ (a), $v_0=0.1$ (b), and $v_0=0.5$ (c). With increasing time delay, the signal spreads slower and the dispersion relation becomes a linear curve overlaid by oscillations. d) The direction-resolved dispersion relations for agents in the upper (positive direction, red line) and the lower (negative direction, yellow line) half-plane for $v_0 = 0.5$ and $\tau=5$. The corresponding overall dispersion relation (total, blue line) oscillates between the direction-resolved dispersion relations. Parameters used: $N=10^3,\ L_\mathrm{F}=20,\ \varphi=0.1,\ N_\mathrm{runs}=10^2$.}
	\label{fig:Dispersion}
\end{figure}

In Fig.~\ref{fig:Dispersion}a-c, we show the main results of this section, which are the dispersion relations for different speeds $v_0$ and time delays $\tau$. They were obtained as follows. We evaluated the times $T_i$ for the maximum signal strength $\dot{\theta}_i(t)$  for all $N N_{\rm runs}$ particles in all $N_{\rm runs}$ simulation runs. Then, we calculated the distances $d_i(T_i)$ from the leader at time $t=0$ to particle $i$. Finally, we averaged the values obtained for the individual particles over  $N_{\rm runs}$ nearest neighbours (the $\left<\bullet \right>_{nn}$ average introduced in Sec.~\ref{Sec Signal}). The relation of both averages is shown in the plot.  

For distances $d_i \approx L_F/2 = 10$ near the boundary of the flock, the dispersion relations are strongly affected by boundary effects, leading to a `flattening' of the curves.
For all considered speeds, a retardation of the interactions slows the signal propagation through the system. In accord with Ref.~\cite{Geiss2021lattice}, the dispersion relations are diffusive ($\left<d_i \right>_{nn} \sim \sqrt{\left<T_i \right>_{nn}}$) for $\tau = 0$ and small speeds $v_0$.  For a positive $v_0$, increasing $\tau$ leads to gradually more linear dispersion relations with slopes of the order of the leader's speed $v_0 \sin \varphi$ relative to the rest of the flock. Specifically, for $v_0=0.1$ and $\varphi = 0.1$ we have $v_0 \sin \varphi\approx 0.01$ and we find \textcolor{black}{ the slope $0.01$ for $\tau = 5$ and $\tau=20$. For $v_0=0.5$ also the dispersion relation for $\tau=0$ has a significant linear part. The slopes corresponding to all measured dispersion relations for this velocity are approximately $0.4$, while $v_0 \sin \varphi\approx 0.5$. The agreement between the velocity of the leader relative to the rest of the flock and the slopes of the dispersion relations strongly suggests that the spreading of information is, in these regimes, predominantly due to the convective transport caused by the motion of the leader and not due to the conductive transport effected by the interparticle interactions within the flock.}

For longer delay times, the dispersion relations exhibit \textcolor{black}{downward spikes periodically occurring with the period $\tau+1$, an overall linear increase of the upper envelope, and a diffusive increase in the lower envelope.} In Fig.~\ref{fig:Dispersion}d, we show that the dispersion relations roughly oscillate between the values attained for specific \textcolor{black}{spreading directions of the signal with respect to the leader.} To demonstrate this, we display dispersion relations calculated from the part of the flock in the upper (positive) and lower (negative) half-planes. For the chosen turning angle $\varphi = 0.1$, the leader moves toward the former and away from the latter. In accord with the findings of Ref.~\cite{Geiss2021lattice}, the dispersion relation for the positive direction is, for large enough speed, linear, while the dispersion relation for the negative direction remains diffusive, regardless of $v_0$ and $\tau$. \textcolor{black}{One can thus conclude that introducing the time delay induces some filtering out of the conductive component of the dispersion relation (the weight given to the conductive component corresponds to the periodically occurring spikes, which involve only a few data points compared to the regions between them, corresponding to the convective spreading). Besides, it was found in Ref.~\cite{holubec2021scaling} that increasing the time delay for a nonzero speed $v_0$ increases the effective speed of the individual agents. As the convective component of the dispersion relation becomes more pronounced with increasing speed, these two effects make the dispersion relation gradually more convective with increasing time delay (except for the spikes).}

\textcolor{black}{Using a simplified spin-wave theory for the (continuous-time) VM, valid for small perturbations around a common direction of the whole flock, the dispersion relation for the delay VM can be obtained analytically. The linearized model is derived, e.g., in the Supplementary Information of Ref.~\cite{holubec2021scaling}. It reads
\begin{multline}
\tilde{\eta}\dot{\varphi}(\mathbf{r},t) = J n_c a^2 {\mathcal{4}} \varphi(\mathbf{r},t-\tau) + J n_c \left[\varphi(\mathbf{r},t-\tau) - \varphi(\mathbf{r},t) \right] \\+ \sqrt{2D a^3}\xi(\mathbf{r},t),
\label{eq:varphiFin}
\end{multline}
where $\varphi(\mathbf{r},t)$ describes fluctuations in orientation of the individual particles. The positive parameters $J$ and $n_c$ determine the strength of the interparticle interactions, $a>0$ denotes a typical interparticle distance, $\tilde{\eta}>0$ sets the intrinsic relaxation timescale, and $D$ determines the amplitude of the Gaussian white noise $\xi(\mathbf{r},t)$. The corresponding dispersion relation can be determined after neglecting the noise and taking the Fourier transform of the linearized dynamical equation in space and time~\cite{cavagna2018physics}. Denoting the Fourier variables corresponding to space and time as $\omega$ and $k$, respectively, and assuming just one spatial dimension, we obtain
the transcendental equation
\begin{equation}
    \tilde{\eta}\omega = -i J n_c \left[(k^2 a^2 - 1)\exp(i\omega\tau) + 1 \right].
		\label{eq:transEQ}
\end{equation}
For $\tau=0$, it yields the purely imaginary dispersion relation $\omega(\mathbf{k}) =  - i J  n_c a^2 k^2$, describing an exponentially damped diffusive transport. For a non-zero delay, 
\begin{equation}
 \omega(\mathbf{k}) =  \frac{i}{\tau} \left\{
 -\tilde{J}
 + W_n\!\left[
(1- a^2k^2)\tilde{J} 
\exp(\tilde{J})
 \right]
 \right\}, 
 \label{eq:omk}
\end{equation}
$n=0,\pm 1, \pm 2, \ldots$, where $W_n(.)$ is the $n$-th branch of the Lambert W function, which is a multivalued function frequently entering solutions to time-delay systems~\cite{geiss2019brownian}, and $ \tilde{J} = J n_c \tau/\tilde{\eta}$. For $n\neq 0$ the dispersion relation has non-zero real components which correspond to a damped a wave-like (ballistic) transport~\cite{cavagna2018physics}.
Unfortunately, all these insights still do not provide a fully satisfactory intuitive explanation of the form of the dispersion relation observed in our simulations. Gaining such an intuition will thus require additional efforts.}

To put these results into a broader context, we note that, from the point of view of natural systems, \textcolor{black}{the spikes} may be considered an artifact of choosing precisely the same delay time for all interactions. In practical applications, the time delay is more likely to be randomly distributed among the individuals, so that \textcolor{black}{the spikes} are expected to average out, at least to some extent. With this in mind, the robust conclusion is that the retarded interactions embodied in the delay VM induce linear dispersion relations, similarly as to what inertia does in a generalized VM with underdamped orientations~\cite{cavagna2018physics}. \textcolor{black}{In the underdamped variant of the VM, the dispersion relation can be derived analytically~\cite{Cavagna2015Silent} and the ballistic/convective transport can be intuitively explained as a result of an additional (approximately) conserved degree of freedom~\cite{cavagna2018physics}. 
}

\section{Linear Response to a leader} \label{sec: linear response}

So far, we have analyzed the response of a  Vicsek flock with retarded interactions in terms of signal propagation, as appropriate in the most general context of its applications. However, in a more physics-focused context, it may be more natural to interpret the results within the framework of linear response theory and the fluctuation-dissipation relations known for equilibrium systems. In the latter case, non-invasive measurements of correlation functions suffice to know the response functions for arbitrary weak perturbations~\cite{zwanzig2001nonequilibrium}. However, for systems far from equilibrium\textcolor{black}{, such as the VM}, such practical relations are generally not available~\cite{Altaner2016,Jung2021} \textcolor{black}{and one has to measure the real response of the system to specific perturbations}.

In accord with our previous discussion, we will investigate the response of the delay VM to perturbations induced by a (or various) leader(s). To provide results independent of initial conditions, we consider perturbations that bring the system from one steady state to another perturbed one. The VM eventually converges to a steady state, which is independent of the initial conditions, for a non-zero rotational noise and if the model is solved, e.g., with periodic boundaries. The collective maneuver scenario of Sec.~\ref{Sec Turns} requires that the size of the periodic box is inversely proportional to the curvature of the leaders' trajectory (so that it fits in the box), which is not a suitable situation for computer analysis. On the other hand, the perturbation scenario of Sec.~\ref{Perturbation} induces a relaxation of the system to its (slightly rotated) initial steady state, and one can measure the characteristics of this relaxation, which have to some extent been discussed in Sec.~\ref{Perturbation}.
We now turn to the scenario of Ref.~\cite{pearce2016linear}, where one applies a permanent torque to the leader(s), which induces a steady state with a nonzero circular flux of the particles.

\begin{figure}
    \centering
    \includegraphics[width=1.05\linewidth]{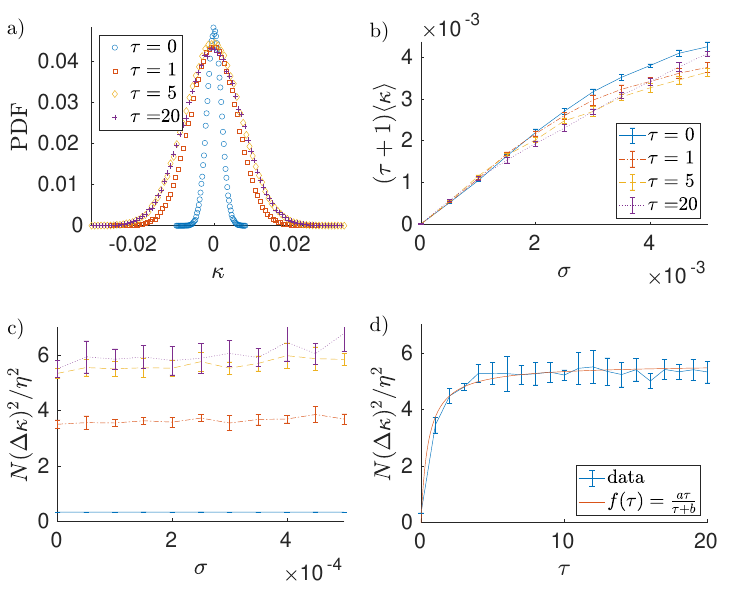}
	\caption{a) The probability distribution of the curvature~\eqref{eq: curvature} for vanishing bias, $\sigma=0$. b) Average curvature as function of $\sigma$ for four delay times. c) Corresponding variance. d) Variance of curvature as function  delay time (symbols with error bars) and the empirical fit $a \tau/(\tau+b)$ with parameters $a = 5.57$ and $b = 0.5$ (solid line). Parameters used: $N=10^3, N_\mathrm{L}=0.05N, \rho_0=4, v_0=0.5$, and $\eta=0.1$.}
	\label{fig:LinearResp}	
\end{figure}

Specifically, we consider a system of $N$ particles inside a square box with sides of length $L$ and periodic boundary conditions. We again assume that the individual particles except for the leader obey Eqs.~\eqref{eq:vtdisc}--\eqref{eq:rtdisc}. However, different from the previous section, we consider a small but nonzero value of the noise $\eta$, such that the unperturbed system eventually reaches a highly ordered \textcolor{black}{(yet noisy)} non-equilibrium steady state.
We initialize the system at time $t=0$ in a polarized state, let it evolve until time $t_I$ when it reaches a steady state \textcolor{black}{independent of the initial data, and then switch} on a perturbation affecting $N_\mathrm{L}$ randomly chosen leaders. Specifically, we assume that velocities of the leaders obey the form  
\begin{equation}
\mathbf{v}_\mathrm{L}(t+1) = v_0 \mathcal{T}_\varphi \mathcal{R}_\eta \Bigl[\mathbf{v}_\mathrm{L}(t) + \sum_{j\neq i } n_{ij}(t-\tau) \mathbf{v}_j(t - \tau) 
\Bigr], \label{eq:vtdisc_leader}
\end{equation}
where the operator $\mathcal{T}_\varphi$ applies a deterministic rotation by an angle $\varphi$ to its argument. Following Ref.~\cite{pearce2016linear}, we study how this perturbation affects the discrete curvature in the overall velocity, defined as
\begin{equation}
	\kappa(t) = \boldsymbol{\bar{\Phi}}(t-1) \times \boldsymbol{\bar{\Phi}}(t), 
	\label{eq: curvature}
\end{equation} 
where $\boldsymbol{\bar{\Phi}}(t)=1/(Nv_0) \sum_i \boldsymbol{v}_i(t)$ denotes the mean orientation of the flock at time $t$. The resulting dependence of the average curvature $\left<\kappa\right>$ on the (relative) strength $\sigma \equiv \varphi N_L/N$ of the perturbation is depicted in Fig.~\ref{fig:LinearResp}. In this section, averages $\left<\kappa\right> \equiv \lim_{t_I\to \infty} t^{-1} \int_{t_I}^{t_I+t} dt' \kappa(t')$ are calculated over time after the system reached the steady state, with $t=2 \times 10^4$ and $t_I = 10^4$. The shown error bars correspond to standard deviations of these averages computed from 10 independent realizations of the numerical experiment.

Fig.~\ref{fig:LinearResp}a illustrates that, for $\sigma=0$, the flock turns randomly and $\kappa(t)$ is a Gaussian random variable with zero mean, $\left<\kappa\right>=0$. Imposing a small directional bias $\sigma \ll \eta$ to the leaders induces an overall non-zero mean curvature, 
\begin{equation}
\langle \kappa \rangle = \alpha \sigma,
\end{equation}
which increases linearly with $\sigma$ (see Fig.~\ref{fig:LinearResp}b).
On the other hand, the variance, $(\Delta \kappa)^2 \equiv \left<\kappa^2\right> - \left<\kappa\right>^2$, in Fig.~\ref{fig:LinearResp}c remains constant.
For our case of highly aligned flocks (small enough $\eta$ and large enough density $N/L^2$), the slope $\alpha$ is approximately given by $(1 + \tau)^{-1}$. 
Notice how the individual curves in Fig.~\ref{fig:LinearResp}b depart from the initially common straight line. The linear-response regime is diminished for both long time delays and large directional biases, and the expression for $\alpha$ holds the better the smaller the product $\sigma \tau$. 
The dependence of the variance on the time delay saturates. Its functional form is well parametrized by $\Delta \kappa^2 \approx a \tau/(\tau+b)$ (see Fig.~\ref{fig:LinearResp}d).
Furthermore, for dense flocks, the variance increases with the noise as $(\Delta \kappa)^2 \propto \eta^2$, \textcolor{black}{independent of the perturbation $\sigma$} (for the corresponding illustrations, see Ref.~\cite{pearce2016linear}). 

In App.~\ref{sec:appendix} we show how this behavior can be understood analytically, based on a simplified model. Here, we provide the emerging intuitive explanation. 
For vanishing delay and in two dimensions, the curvature~\eqref{eq: curvature} can be rewritten as $\kappa(t) = \sin \langle \Delta \theta \rangle_N$ with $\langle \Delta \theta \rangle_N = N^{-1}\sum_i [\theta_i(t)-\theta_i(t-1)]$ denoting the mean change of orientation during one time step. For small perturbations $\varphi$, $\langle \Delta \theta \rangle_N\ll 1$ and thus $\kappa \approx \langle \Delta \theta \rangle_N$. For weak noise, high average density, and vanishing delay, the local density of the VM is approximately homogeneous~\cite{cavagna2021equilibrium}. As a result, the total orientation (or information) $N^{-1}\sum_i \theta_i(t)$ is for $\varphi = 0$ approximately conserved \cite{Geiss2021lattice}. For $\varphi > 0$, it is then reasonable to assume that the change in the total orientation is directly given by the directional bias applied to the leaders, i.e., $\langle \Delta \theta \rangle_N = 1/N \sum_{1}^{N_\mathrm{L}}\varphi=\sigma$. 
For a nonzero delay, the orientational change is \textcolor{black}{stretched out} over $\tau+1$ time steps, reducing $\langle \Delta \theta \rangle_N$ by the factor $\tau + 1$. Fig.~\ref{fig:LinearResp}b can thus be regarded as a verification of the (approximate) information conservation in a highly ordered Vicsek flock. 
Large biases $\varphi$ can break the highly ordered state and introduce growing density fluctuations which cause the breakdown of orientation conservation~\cite{Geiss2021lattice} and, as a consequence, linear response. Similarly, as we have demonstrated above, long delays tend to decouple the leaders from their neighborhoods\textcolor{black}{, which also induces fluctuations in density, with the same consequences.}

In this vein, the variance of the curvature can be approximated as
\begin{equation}
	\Delta(\kappa)^2 = \Delta\left(\frac{1}{N}\mbox{$\sum_i$} \kappa_i \right)^2 = \frac{1}{N} \Delta(\kappa_i)^2,
\end{equation} 
where $\kappa_i = \Delta \theta_i$. To expand the square, we assumed that the orientational fluctuations $\Delta \theta_i$ of individual agents are statistically independent, which is sensible for small perturbations $\varphi$ of a highly aligned state. \textcolor{black}{Assuming that the fluctuations are caused solely by the noise, we get $\langle \Delta\theta_i^2 \rangle \approx \int_{-\eta}^{\eta} \tilde{\eta}^2 \mathrm{d}\tilde{\eta}/(2\eta) = \eta^2/3$.} Altogether, we thus find the variance $\Delta(\kappa)^2 \propto \eta^2/N$, which explains the findings in Ref.~\cite{pearce2016linear}. According to our argument, this scaling should be a direct consequence of the law of large numbers. Introducing a finite delay enhances effects of the noise on the dynamics and triggers oscillations~\cite{geiss2019brownian}. As a result, the curvature variance strongly increases with the delay. Unfortunately we were not able to rationalize the observed quantitative dependence of the variance on the time delay, shown in Fig.~\ref{fig:LinearResp}d, from the equations of motion. 


\section{Conclusion and outlook}\label{Sec: Conclusion}

In summary, we have studied the response to orientation perturbations in the 2-dimensional delay Vicsek model. We have found that a time delay diminishes the ability of strongly aligned systems to follow a leader moving along a predetermined path (different from \textcolor{black}{what has been reported for inertia}~\cite{cavagna2015flocking}) and increases the stability of these systems \textcolor{black}{against} sudden local fluctuations of orientation. Both effects are caused by the decreased information (orientation) propagation through the system and the \textcolor{black}{stretched} time-correlations in the system, due to the delay. Their combination explains the findings in Refs.~\cite{piwowarczyk2019influence,holubec2021scaling} that short delays \textcolor{black}{facilitate} and long delays hinder flocking in the delay VM. 

In the scenario of a sudden turning pulse applied to a leader, the source of information or perturbation is well localized in space and time.   Inspired by~\cite{cavagna2018physics}, we used the induced orientational
acceleration as a measure for the signal strength to study the propagation of information (orientation)  through the system. As typical for delay systems~\cite{geiss2019brownian}, a time delay causes oscillations. In the present context, they are manifestly seen in the orientational accelerations. Their amplitude, period, and decay time (and thus system memory) all increase with growing delay~\cite{holubec2021scaling}. To measure the speed of information propagation through the system, we used the maximum of the acceleration pulse transmitted through the flock. We found that the dispersion relation for the signal spreading is diffusive (the distance traveled by the signal $\propto$ square root of the traveled time) for small speeds and short time delays, but becomes increasingly ballistic (linear) for long delays and high speeds. 

We have \textcolor{black}{finally} investigated the linear response of highly ordered systems to a weak orientational bias, applied to a subgroup of agents (leaders). We  found that the overall change of the flock's orientation is a linear function of the perturbation for short enough time delays and weak enough orientational biases. The linear response regime coincides with the regime of low density fluctuations, where the total information (or average orientation) of the flock is (approximately) conserved. Furthermore, we provided intuitive and analytical arguments for the observed functional dependencies of the mean total orientational change and the corresponding variance.
 
While our work provides a first understanding of retardation effects in active many-body dynamics, it also raises some further questions. The choice of a universal value of the discrete time delay for all members of the flock certainly oversimplifies the conditions prevailing in natural flocks and swarms, and it would be desirable to get a better understanding \textcolor{black}{of} the concomitant internal averaging and its role in suppressing the oscillatory instabilities observed in our idealized model.  Also, the reaction capabilities of natural agents will  usually be described by more complicated memory functions~\cite{geiss2019brownian,Loos2021}. Further, it could be interesting to investigate more thoroughly the effects of time delay on the collective memory of the flock, which is an important ingredient in the understanding of so-called swarm intelligence~\cite{bonabeau1999swarm,brambilla2013swarm}.  Our present results confirm the natural expectation that the system's relaxation time is sensitive to the delay time.
Another direction could be to study information spreading in more complex interaction networks, e.g., with distributed individual  weights for the flock members. Such leadership hierarchies are often observed in  natural groups~\cite{chase1974models,nagy2010hierarchical}, and it has been argued that they increase the group fitness~\cite{nagy2010hierarchical,shen2008cucker,balazs2020adaptive}. For instance, it is known that a few (well-connected) agents, known as superspreaders, lead to a rapid increase in disease (or fad) spreading  in epidemiology~\cite{olinky2004unexpected,meyers2005network,grossmann2021ode}.

\section*{Acknowledgments}
We acknowledge funding through
a DFG-GACR cooperation by the Deutsche Forschungsgemeinschaft (DFG-Code KR 3381/6-1) and by
the Czech Science Foundation (GACR Project No 20-02955J). VH was supported by the Humboldt Foundation. DG acknowledges funding by International
Max Planck Research Schools (IMPRS) as well as by Deutscher Akademischer Austauschdienst (DAAD).

\appendix

\section{Exactly solvable model}
\label{sec:appendix}

The dependence of the first two moments of the mean curvature~\eqref{eq: curvature} on the delay time $\tau$, discussed in Sec.~\ref{sec: linear response}, can be obtained using a simple mean-field model. Consider a system of $N$ mutually interacting particles with a single leader (index $\mathrm{L}$) that interacts with the average flock orientation $\theta_0(t)$ of the other $N-1$ flock members. The orientations obey the dynamical equations
\begin{equation}
	\theta_\mathrm{L}(t+1) = \frac{1}{N} \left[ \theta_\mathrm{L}(t) + (N-1) \theta_0(t-\tau) \right] + \varphi + \xi_\mathrm{L}(t), 
	\label{eq: analyticalModel1}
\end{equation}
\begin{multline}	
	\theta_0(t+1) = \frac{1}{N} \left[ \theta_0(t) + (N-2) \theta_0(t-\tau) \right. \\
 \left. + \theta_\mathrm{L}(t-\tau) \right] + \xi_0(t).
	\label{eq: analyticalModel}
\end{multline}
The symbols $\xi_\mathrm{L}(t)$ and $\xi_0(t)$ denote independent $\delta$-correlated angular noises uniformly distributed within the interval $[-\eta,\eta]$. The mean orientation of the system, $O(t)\equiv\langle \theta_i(t) \rangle_{N} = \left[ \theta_\mathrm{L}(t)+(N-1)\theta_0(t) \right]/N$, obeys the equation
\begin{equation}
    O(t+1) = \frac{1}{N} \left[ O(t) + (N-1) O(t-\tau) \right] + \sigma + \hat{\xi}(t)  \label{eq:MeanOrientation}
\end{equation}
with $\sigma =\varphi/N$. The noise $\hat{\xi}(t)=[\xi_\mathrm{L}(t)+(N-1)\xi_0(t)]/N$ has variance $\langle\hat{\xi}^2\rangle = (N^2-2N+2)\eta^2/(3N^2)$. 

The difference equation~\eqref{eq:MeanOrientation} for the column vector $\boldsymbol{O}(t) \equiv (O(t),\dots,O(t-\tau))^\intercal$  can be solved using its matrix representation,
\begin{equation}
    \boldsymbol{O}(t+1) = M \boldsymbol{O}(t) + \boldsymbol{\sigma} + \boldsymbol{\hat{\xi}}(t).
\end{equation}
Here, $\boldsymbol{\sigma} \equiv (\sigma,0,\dots,0)^\intercal$ and $\boldsymbol{\hat{\xi}}(t) \equiv (\hat{\xi}(t) ,0,\dots,0)^\intercal$ are column vectors of length $\tau+1$, and $M$ is a $(\tau+1) \times (\tau+1)$ square matrix with all elements equal to zero except for $M_{11} = 1/N$, $M_{1(\tau + 1)} = (N-1)/N$, and $M_{(i+1)i} = 1$, $i = 1,\dots,\tau$. 
The formal solution to this equation is
\begin{equation}
  \boldsymbol{O}(t) = M^{t} \boldsymbol{O}(0) + \sum_{i=0}^{t-1} M^i [\boldsymbol{\sigma} + \boldsymbol{\hat{\xi}}(t-i-1)],
\end{equation}
where $\boldsymbol{O}(0)$ is the initial condition, which we set to 0 (all particles aligned with the $x$-axis). Our goal is to calculate the change $\kappa(t) = O(t)- O(t-1)$ of the mean orientation. The corresponding vector $\boldsymbol{\kappa}(t) = (\kappa(t),\dots,\kappa(t-\tau))^\intercal$ is given by $\boldsymbol{O}(t) - \boldsymbol{O}(t-1)$ and reads
\begin{equation}
  \boldsymbol{\kappa}(t) =  M^{t-1} [\boldsymbol{\sigma} + \boldsymbol{\hat{\xi}}(0)] + \sum_{i=0}^{t-2}M^i [\boldsymbol{\hat{\xi}}(t-i-1)-\boldsymbol{\hat{\xi}}(t-i-2)].
  \label{eq:vectorxi}
\end{equation}
At late times, $\kappa(t)$ reaches a steady state, in which all elements of the vector $\boldsymbol{\kappa}(t)$ have the same mean and variance, given by $\langle\kappa\rangle$ and $\langle\Delta \kappa^2\rangle$. Specifically, we find
\begin{eqnarray}
\langle\boldsymbol{\kappa}\rangle &=& M_\infty \boldsymbol{\sigma},\\
\langle\boldsymbol{\kappa}^2\rangle - \langle\boldsymbol{\kappa}\rangle^2 &=&
\langle\hat{\xi}^2\rangle R_{11},
\label{eq:2}
\end{eqnarray}
where $M_\infty \equiv \lim_{t\to\infty} M^t$, $R = M_\infty^\intercal M_\infty + 2 S - M_\infty^\intercal S - S M_\infty$, and $S \equiv  \sum_{i=0}^\infty (M^i)^\intercal M^i$. The right hand side of Eq.~\eqref{eq:2} results from multiplying the sum in Eq.~\eqref{eq:vectorxi} by itself and using the Markov condition $\langle\hat{\xi}(t)\hat{\xi}(t')\rangle = \langle\hat{\xi}^2\rangle \delta_{t t'}$ for the stationary noise correlation function. 

\begin{figure}[t]
    \centering
    \includegraphics[width=1.05\linewidth]{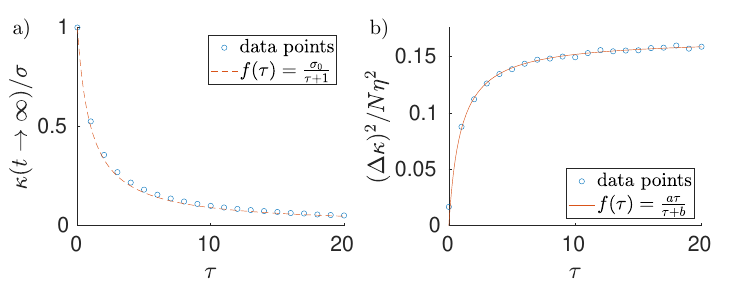}
	\caption{Noise-averaged stationary orientational change. Symbols depict the mean (a) and variance (b) as functions of the delay time. The dashed line in a) is the analytical prediction~\eqref{eq:anal}, and the dashed line in b) is the empirical fit $a\tau/(\tau+b)$ with parameters $a = 0.17$ and $b = 0.93$.}
	\label{fig:AnalyticalModel}
\end{figure}

It is straightforward to verify, e.g., by calculating left and right eigenvectors corresponding to the eigenvalue 1, that the stationary matrix $M_\infty$ has all elements in its first column equal to $c (N-1)/N$, with $c = [(N-1)/N + \tau]^{-1}$. Its remaining elements are equal to $c$. Hence, we find that all elements of $\left<\boldsymbol{\kappa}\right>$ are given by $\sigma c$, and thus, in agreement with the discussion in Sec.~\ref{sec: linear response},
\begin{equation}
 \left<\kappa\right> = \frac{\sigma}{(N-1)/N + \tau}.
 \label{eq:anal}
\end{equation}
In Fig.~\ref{fig:AnalyticalModel}a, we show agreement of this prediction with a numerical solution to Eqs.~\eqref{eq: analyticalModel1} and \eqref{eq: analyticalModel}.

Equation~\eqref{eq:2} is a scalar product of two identical vectors of length $(\tau+1)$ containing in all entries the variance $(\Delta\kappa)^2$. Thus we find
\begin{equation}
  (\Delta\kappa)^2 = \frac{N^2-2N+2}{N^2}\frac{\eta^2}{3(\tau + 1)} R_{11}. 
\end{equation}
While we were not able to extract the $\tau$- and $N$-dependence of the matrix element $R_{11}$ analytically, Fig.~\ref{fig:AnalyticalModel}b shows that the $\tau$-dependence of the calculated $(\kappa)^2$ is well described by the empirical formula $a\tau/(\tau + b)$, already employed in Sec.~\ref{sec: linear response}.

\bibliography{bibfile}

\end{document}